\documentclass[]{aa}
\usepackage{txfonts}
\usepackage{graphicx}
\usepackage{aalongtable}

\def\gam1leo{$\gamma^1$\,Leo }
\def\tauceti{$\tau$ Ceti}
\def\Te  {$T_{\rm eff}$}
\def\lgg {$\log g$}

\def\Msun{$M_{\sun}$}
\def\mps {$\rm\, m\,s^{-1}$}
\def\kmps {$\rm\, km\,s^{-1}$}
\def\angstrom {\AA}

\begin{document}
\title{Detection of a Planetary Companion around the giant star \gam1leo}
  \thanks{Based on observations made with the BOES spectrograph at the 1.8 m telescope of BOAO, KASI}
  \fnmsep
  \thanks{Tables 3 is only available in electronic form at the CDS via anonymous ftp to cdsarc.u-strasbg.fr (130.79.128.5)
          or via http://cdsweb.u-strasbg.fr/cgi-bin/qcat?J/A+A/}
\author
 {Inwoo Han \inst{1}
 \and B.~C. Lee \inst{1, 2}
 \and K.~M. Kim \inst{1}
 \and D.~E. Mkrtichian \inst{3, 4}
 \and A.~P. Hatzes \inst{5}
 \and G. Valyavin \inst{6}
 }

\institute
 {
  Korea Astronomy and Space Science Institute, 61-1, Whaam-dong, Yuseong-gu, Daejeon, 305-348, Korea
   \and
  Department of Astronomy and Atmospheric Sciences, Kyungpook National University, Daegu, 702-701, Korea
   \and
  Astrophysical Research Center for the Structure and Evolution of the Cosmos, Sejong University, Seoul, 143-747, Korea
   \and
  Crimean Astrophysical Observatory, Nauchny, Crimea, 98409, Ukraine 
   \and
  Th\"uringer Landessternwarte Tautenburg, Sternwarte 5, D-07778, Tautenburg, Germany
   \and
  Observatorio Astron\'{o}mico Nacional SPM, Instituto de Astronom\'{i}a, Universidad Nacional Aut\'{o}noma
  de M\'{e}xico, Ensenada, BC, M\'{e}xico
 }
\date{Received 22 May 2009 / Accepted 13 October 2009}

\abstract
{}
{
Our primary goal is to search for planets around intermediate mass stars.
We are also interested in studying the nature of radial velocity (RV) variations of K giant stars.
}
{
We selected about 55 early K giant (K0 - K4) stars brighter than fifth magnitude that were observed  using BOES,
a high resolution spectrograph attached to
the 1.8 m telescope at BOAO (Bohyunsan Optical Astronomy Observatory).
BOES is equipped with $I_2$ absorption cell for high precision RV measurements.
}
{
We detected a periodic radial velocity variations in the K0 III star \gam1leo with a period of P = 429 days.
An orbital fit of the observed RVs yields a period of P = 429 days, a semi-amplitude of K = 208 \mps, and an eccentricity of $e = 0.14.$
To investigate the nature of the RV variations, we analyzed the photometric, Ca II $\lambda$ 8662 equivalent width,
and line-bisector variations of \gam1leo.
We conclude  that the detected RV variations can be best explained by a planetary companion with
an estimated mass of m $\sin i = 8.78 M_{Jupiter}$ and a semi-major axis of $a = 1.19$ AU, assuming a stellar mass of 1.23 \Msun.
}
{}


\maketitle

\section{Introduction}
Since the first exoplanet around a main sequence star was discovered in 1995, more than 300 exoplanets have been detected (http://exoplanet.eu/catalog.php).
Among them, the majority of exoplanets candidates has been detected by using radial velocity methods around
late F-G and K main-sequence stars.

The existence of planets around intermediate mass early-type stars and their planetary parameters have
not been investigated well due to their fast stellar rotational velocities and the lack of an appropriate number of sharp
absorption lines in the stellar spectra. When intermediate mass stars evolve toward the red giant stage, they go
through the G and K-giant phase where many sharp absorption lines appropriate for high precision RV measurements
are available. Therefore, G and K giant stars are suitable targets for detecting exoplanets with the RV technique.
There are several ongoing exoplanet survey projects around giant stars
(Setiawan et al. 2005, Hatzes et al. 2005, Sato et al. 2007, Johnson et al. 2007,
Lovis \& Mayor 2007, Niedzielski et al. 2007, and Liu et al. 2008).
Now, more than 20 exoplanets have been detected around giant stars and from
this sample some statistical studies on the planetary systems around giants have been made
(Pasquini et al. 2007, Hekker et al. 2008, and Sato et al. 2008).

We started our precise RV survey using the 1.8m telescope at BOAO (Bohyunsan Optical Astronomy Observatory) in 2003
in order to search for exoplanets and to study the pulsations and stellar surface activity of K-giant stars.
Our sample consists of 55 K0 - K4  giant  stars.
Most of them are brighter than fifth magnitude. Results from  the first 3 years of
our survey show that the majority (more than 90\%) of our sample of K-giants show RV dariations.
Among them, we found spectroscopic binaries, pulsating variables (see Kim et al. 2006), and several periodic RV variable stars.
Our survey confirmed an exoplanet around $\beta$ Gem (Han et al. 2008) discovered by Hatzes et al. (2006) and Reffert et al. (2006).
In this paper we present a new exoplanet detection around the giant star \gam1leo.

\section{Observations and data reduction}
The RV observations of \gam1leo have been carried out from May 2003 until May 2009 using the fiber-fed
high resolution Echelle spectrograph BOES (Kim et al. 2007), attached to the 1.8 m telescope at BOAO.
BOES has five fibers with diameters of 80, 100, 150, 200, and 300
$\mu$m. The measured resolving power for each fiber is $R$ = 90,000, 75,000, 60,000,
45,000, and 30,000 respectively. Using a 2k x 4k CCD, the wavelength coverage of BOES is 3600 -- 10500
\,{\AA} with 86 spectral orders in a single exposure. BOES is also equipped
with $I_2$ absorption cell for precise radial velocity measurements.
All observations of \gam1leo were conducted by using 80$\mu$m fiber to
achieve the highest resolving power and RV accuracy. The typical exposure
time was 200 s to give a S/N ratio of about 150.

The extraction of the normalized 1D spectra was carried out using IRAF software package
by following standard procedures (bias subtraction, background removal, flat fielding, wavelength calibration).
After extracting normalized 1D spectra, the RVs were computed using a code called
RVI2CELL which was developed at BOAO by Han et. al (2007).

\section{The target star}

\begin{table}
\label{targettable}
\begin{center}
\caption{some stellar parameters of \gam1leo}
\begin{tabular}{ccl}
\hline
parameter & value & reference \\
\hline
$m_{V}$           & 2.01             & Hipparcos   \\
$B - V$           & 1.13             & Hipparcos   \\
parallax (mas)    & 25.96 $\pm$ 0.83 & Hipparcos \\
$M_{V}$           & $-0.92 \pm 0.069$& derived   \\
diameter (mas)    & 7.7   $\pm$ 0.3  & Dyck et al. (1998) \\
radius($R_\odot$) & 31.88 $\pm$ 1.61 & derived \\
mass  ($M_\odot$) & 1.23  $\pm$ 0.21 & This study  \\
$v \sin i$ (km/s) & 1.1   $\pm$ 1.3  & de Medeiros \& Mayor(1999) \\
                  & 2.6              & Gray (1982) \\
                  & 3.9              & Massarotti et al.(2008)  \\
                  & 2.7              & this study  \\
\Te               & 4300             & Tomkin et al. (1975)        \\
                  & 4470             & McWilliam (1990)   \\
                  & 4330 $\pm$ 15    & This study \\
\lgg              & 1.7              & Tomkin et al. (1975)        \\
                  & 2.35             & McWilliam (1990)   \\
                  & 1.59             & This study \\
\rm{[Fe/H]}       & -0.49 $\pm$ 0.12 & Mcwilliam (1990)   \\
                  & -0.51 $\pm$ 0.05 & This study \\
$\zeta$(km/s)     & 2.4              & Mcwilliam (1990)   \\
                  & 1.5              & This study \\
\hline
\end{tabular}
\end{center}
\end{table}

\gam1leo (HIP50583, HD 89484, HR 4057, Al Gieba) is the bright (V=2.01, $B-V$ = 1.13, K0III)
component of a visual binary system (ADS 7724AB, WDS10200 +1950, STF 1424AB).
The Hipparcos parallax is 25.96 $\pm$ 0.83 mas, which results in a  distance of 38.5 pc.
The double star nature of this system was first discovered by William Herschel in 1792.
Although the orbit is still not well determined, Mason et al. (2006) listed the orbital elements as
P = 510 years, e = 0.84, $\alpha = 4.24$ arcsec, and i = 76$^{\circ}$.
At present the separation between the two stellar components is about 4.6 arcsec.

The stellar parameters of \gam1leo has been determined by several authors.
Table 1 summarizes the results.
We determined the atmospheric parameters of \gam1leo directly from the spectra
obtained using the BOES spectrograph without the iodine cell.
By using 247 neutral and 11 ionized iron absorption lines and measuring their equivalent widths,
we estimated the \Te, \lgg, $v_{t}$, and [Fe/H] of the star using the program TGVIT by Takeda et al. (2005).
We also estimated the projected rotational velocity $v \sin i$ using the procedure given by Takeda et al. (2008).
To estimate the mass of the star, we used the online evolutionary track tool from da Silva et al. (2006)
and Girardi et al. (2000).
This resulted in  $M = 1.23\pm 0.21 \,M_\odot $ and \lgg\, = 1.49 $\pm$ 0.11.
We note that the \lgg\, value estimated by using the evolutionary track is in good agreement with
the result of 1.59 from our spectroscopic estimation.

\section{The Observed Radial Velocities and Orbital Analysis}
\begin{figure}
\label{fig_orb}
\centering
\includegraphics[width=9cm]{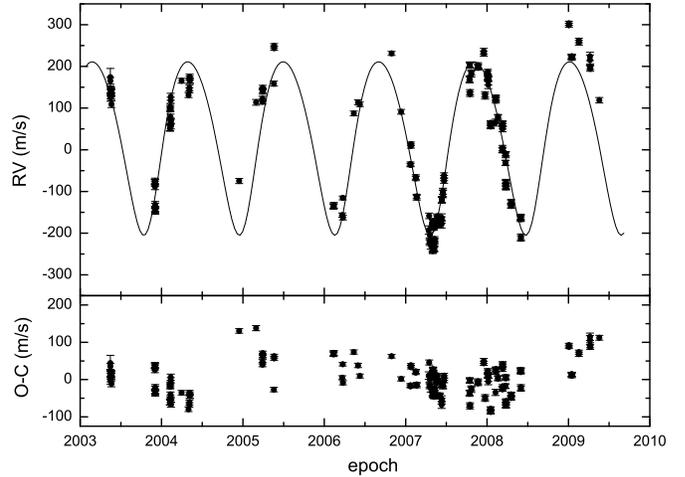}
\caption[]{ Top: The observed radial velocities of \gam1leo.
The orbital fit is shown by the solid line.
Bottom: Residual RV after subtracting the orbital fit. The rms of the residuals is
40 ms$^{-1}$.
}
\end{figure}

The RV measurements obtained from May 2003 until May 2009 are shown in Figure 1 and listed in Table 3.
(Table 3 is shown only in a machine-readable form in the online journal.)
The typical internal error of the RV measurements is about 5 to 7$\rm\, m\,s^{-1}$.
To check the long term RV measurements accuracy of BOES, we have observed a RV standard star \tauceti.
Figure 2 shows the RV variation of \tauceti. From the result of $\tau$ Ceti monitoring,
we estimate the RV measurements accuracy of BOES to be about 8 \mps.

\begin{figure}
\label{tai_ceti}
\centering
\includegraphics[width=9cm]{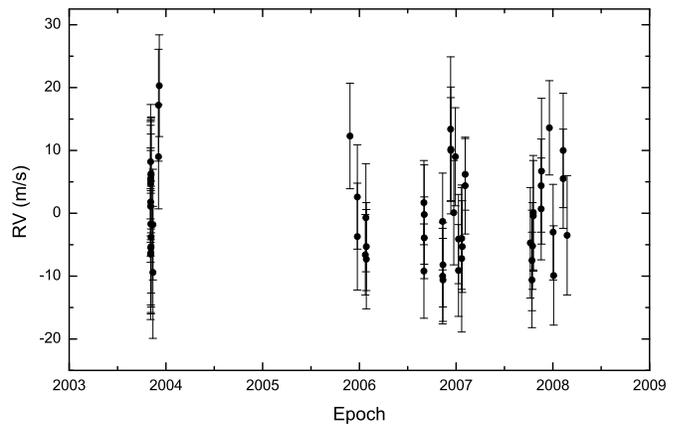}
	\caption[]{Variation of the RV of standard star {\tauceti} observed by BOES.
The rms scatter of the observed RV is 7.5 \mps.
}
\end{figure}

From the observed RV data, we can see a periodicity immediately.
An analysis using a Lomb-Scargle periodogram (Lomb 1976, Scargle 1982) confirmed
a presence of strong power at the frequency f = 0.0023 c/d (P = 435 days).
Figure 3 shows the Lomb-Scargle periodogram and window function.
We found that the false alarm probability (FAP) of this periodicity estimated by a bootstrap randomization process is less than $10^{-6}$.
A more detailed analysis by using the non-linear least squares orbital fitting yields
a period of P = 429 days, and a semi-amplitude of K = 208 \mps.
The orbital parameters with their errors are listed in Table 2.

As can be seen from Figure 1, the rms of the RV residuals is 43 \mps
which is significantly larger than the typical RV measurements error of 5 to 7 \mps.
We also noticed some systematic pattern in the residual RV plot shown in the lower panel of Figure 1.
The periodogram analysis of the residuals RVs shows a strong power around P = 1340 days with a FAP less than $0.5 \times$ 10$^{-5}$.
Figure 4 shows the Lomb-Scargle periodogram and window function of the RV residuals.
Although it is still premature to firmly establish the reality and cause of 1340-d periodicity,
the orbital fit yields a semi-amplitude of K = 35\mps, and an eccentricity e = 0.13.
The estimated mass and semimajor axis of the hypothetical companion is $a$ = 2.6 AU, and m $\sin i = 2.14\,M_{Jupiter}$.
The 1340-d period may arise from stellar rotational modulation.
Taking the $v \sin i$ of de Medeiros \& Mayor in Table 1
and the derived stellar radius of 31.88 $R_\odot$ results in a rotational period of $P/\sin i = 1474$ days,
consistent with the residual RV period given the uncertainty in the $v \sin i$ measurement.

\begin{figure}
\label{power_window}
\centering
\includegraphics[width=9cm]{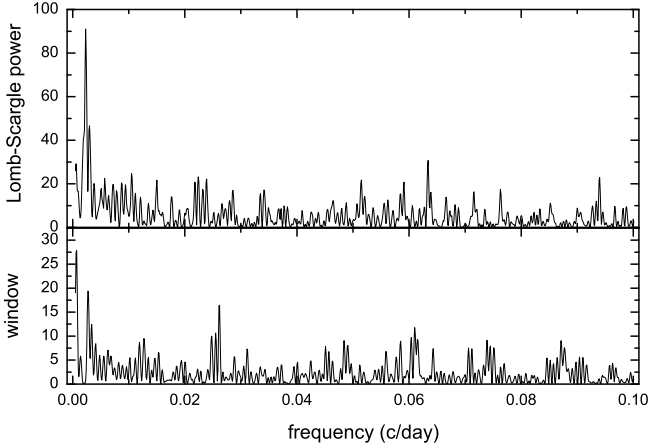}
\caption[]{The periodogram (top) and window function (bottom) of the observed RVs.
 The peak of the periodogram is located at f = 0.0023 c/d or P = 435 days.}
\end{figure}

\begin{table}
\label{tab_orb}
\begin{center}
\caption{Orbital parameters of \gam1leo}
\begin{tabular}{cc}
\hline \hline
Period (days)              & 428.5     $\pm$ 1.25  \\
K ($\rm\, m\,s^{-1}$)      & 208.3     $\pm$ 4.3   \\
e                          & 0.144     $\pm$ 0.046 \\
$T_{periastron}$           & 2451236   $\pm$ 13.5  \\
$\omega$ (deg)             & 206.7     $\pm$ 9.2   \\
rms (\mps)                 & 43                    \\
reduced $\chi^2$           & 48                    \\
m $\sin i$ $(M_{Jupiter})$ & 8.78      $\pm$ 1.0   \\
$a$ (AU)                   & 1.19      $\pm$ 0.02  \\
\hline
\end{tabular}
\end{center}
\end{table}

In addition to 1340-d period, the periodogram in Figure 4 also shows a strong power at P = 8.5 days with an amplitude of about 20 \mps.
The FAP of this periodicity is less than 0.5 $\times$ 10$^{-5}$.
By using the formula in Cox et al. (1972), we calculated the expected fundamental radial pulsation frequency of 8.4 days which is very close to the observed \textbf{8.5-d periodicity}.
If the 8.5-d periodicity is truly due to pulsations, it indicates that our mass and radius estimate for $\gamma^1$ Leo may be reasonable.
More data taken over a shorter time span and with better sampling is needed to confirm the 8.5-d period.

\begin{figure}
\label{residusl_power_window}
\centering
\includegraphics[width=9cm]{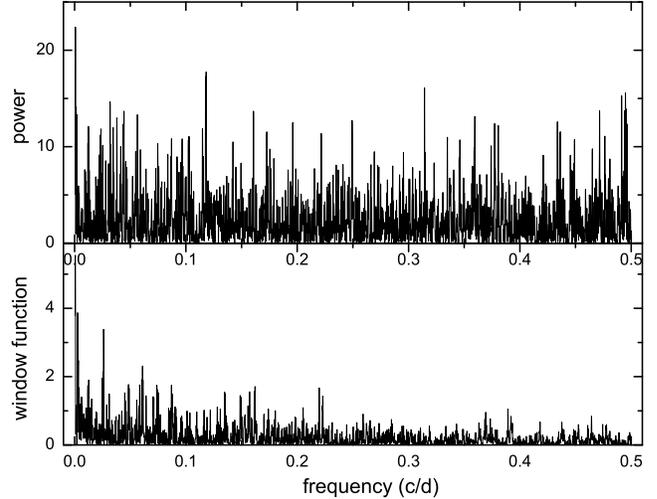}
\caption[]{The Lomb-Scargle periodogram (top) and window function (bottom) of the RV residuals.}
\end{figure}

\section{The cause of the RV variation}
Although the observed 429-d periodic RV variation can be explained by an orbital motion of an unseen companion,
intrinsic mechanisms such as pulsations and rotational modulation by surface features
may cause the RV variations.
To confirm whether the RV variations are due to intrinsic stellar activity,
we investigated the photometric and spectral line profile variations of \gam1leo.

In the next three subsections we will check the intrinsic stellar activity hypothesis by
a) photometric method, b) Ca II $\lambda$ 8662 equivalent width measurements, and
c) the spectral line bisector method.

\subsection{The Hipparcos photometry}

\begin{figure}
\label{hippo_phnpo}
\centering
\includegraphics[width=9cm]{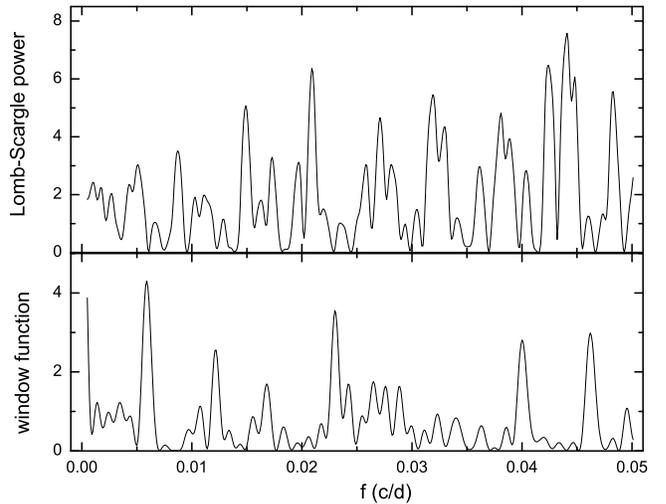}
\caption[]
{The Lomb-Scargle periodogram (top) and window function (bottom) of the Hipparcos photometry of \gam1leo.}
\end{figure}

The photometric database from the Hipparcos satellite contains 50 measurements made over a three year interval.
The rms scatter of the data, after omitting four outliers, is 4.3 mmag.
To see any periodic photometric variation, we performed a periodogram analysis.
Figure 5 shows the Lomb-Scargle periodogram and window function of the Hipparcos photometry.
We did not detect any power around the 429-d orbital period.

If we assume that the rotational modulation of cool spots causes the observed RV variation,
we can estimate the expected photometric variation.
From Table 1, let us assume a mean value of the projected rotational velocity of
$v \sin i = $2.6\kmps.
The this results in a maximum rotational period of $P/\sin i = 620$ days,
which is compatible with the observed 429-d RV variation.
So we may suspect that stellar activity like  cool spots could be  the cause of the observed 429-d RV variation.
On the other hand, in order for spots to produce the observed RV amplitude of the 429-day period a spot filling
factor of  about 0.1 is required, according to the relations in Saar \& Donahue (1997),
Hatzes (2002), and Desort et al. (2007).
Note that if the $v\sin i$ is as low as 1.1 km/s the filling factor will be about a factor of 2 higher.
But the very stable Hipparcos photometry excludes this large filling factor.
It is thus unlikely that cool spot causes the observed RV variations,
although the Hipparcos data were taken at different epoch from our RV observation.

\subsection{The variation of the Ca II $\lambda$ 8662 Equivalent width}
The EW of Ca II $\lambda$ 8662\,\angstrom\, line has been used as an indicator of stellar activity.
In fact, Walker et al. (1992) found the EW variation of this line from the spectra of $\gamma$ Cep
and concluded that the periodic RV variation of $\gamma$ Cep was not due to the orbital motion
by the unseen sub-stellar companion.
Subsequently, Hatzes et al. (2003) showed that there
were no Ca II variations with the RV period which confirmed the planet hypothesis.
We measured the equivalent width of Ca II $\lambda$ 8662 to check any correlation with the 429-d periodic RV variation.
The EW was measured at the central part - from 8661.0 to 8663.5\,\angstrom\, of the line.
The result of our EW measurements is a mean value = 1778.2 and $\sigma$ =  0.7\,m\angstrom.
It implies that the EW of Ca II $\lambda$ 8662 of \gam1leo is very stable.
Figure 6 is the plot of the periodogram of the EW variations.
Figure 6 shows that the EW does not exhibit any significant periodicity at the 429-d RV period.

\begin{figure}
\label{CaII_rvnpower}
\centering
\includegraphics[width=9cm]{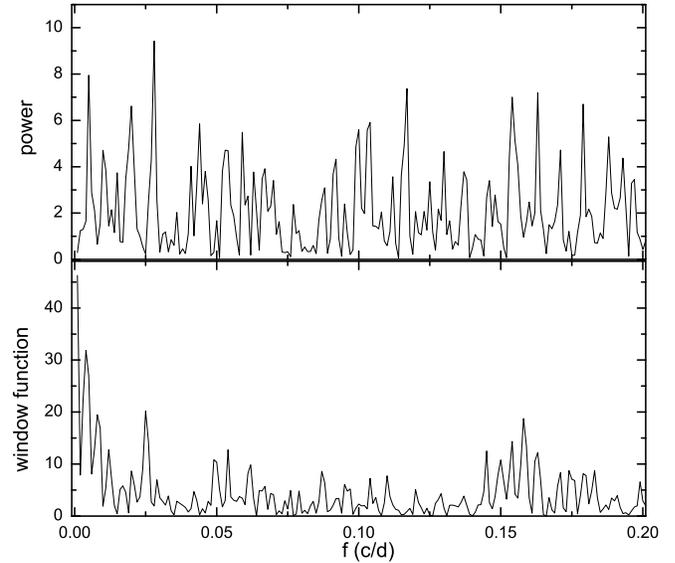}
\caption[]
{The Lomb-Scargle periodogram (top) and window function (bottom) of Ca II $\lambda$ 8662 line EW variation.}
\end{figure}

\subsection{The line bisector variations}
Spectral line bisector analysis has been extensively used to look for line profile shape variations
due to surface features or non-radial pulsations of exoplanet candidate stars.
For our line bisector analysis, unfortunately our spectrograph BOES showed an unstable instrument profile
variation due to the imaging camera optical axis mismatch.
It is a more serious problem at the edge of each order of the spectra.
For RV measurements most of the change in the instrumental profile is taken into account in the modeling process
so this does not pose a serious problem.
For line bisector analysis, instrumental line profile variations directly affect the bisector measurements.
Nevertheless,  we tried the line bisector analysis by selecting the lines very carefully.
We looked for well-isolated and strong lines with central depth larger than 0.6.
Because of the instrument line profile variation,
we tried to find the lines at the central region of the spectral order.
Finally we chose six lines for our bisector analysis
- Ca I 6499.6, Ni I 6643.7, Fe I 6750.2, Ni I 6767.8, Fe I 7780.6, Ni I 7788.9.
As usual, we measured both BVS (Bisector Velocity Span) and BVC (Bisector Velocity Curvature).
BVS is the bisector difference between two different flux levels.
BVC is the difference between two different BVS estimated at different flux levels.
We used three flux levels of 0.8, 0.6, and 0.4 relative to the central line depth to estimate BVS and BVC.

Figure 7 shows the BVS and BVC of all the lines as a function of RV. We do not see any correlation between BVS, BVC and RV.
We computed a periodogram for the BVS and BVC of each line.
We do not see any significant power at P = 429 days from the periodograms.
Figure 8 shows an example of the periodogram obtained from Ni I 6643.7.
Thus our line bisector analysis does not show any evidence that the stellar activity is the cause
of the 429-d peridocitiy in the observed RVs.

\begin{figure}
\label{bvs_rv}
\centering
\includegraphics[width=9cm]{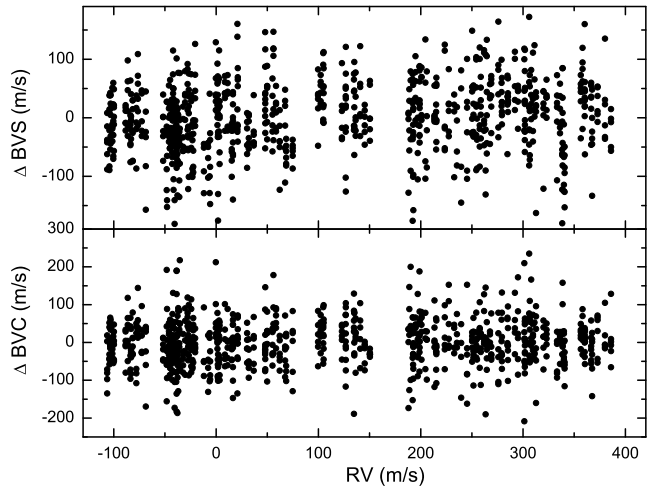}
\caption[]{BVS (top) and BVC (bottom) plotted as a function of RV}
\end{figure}

\begin{figure}
\label{bsa}
\centering
\includegraphics[width=9cm]{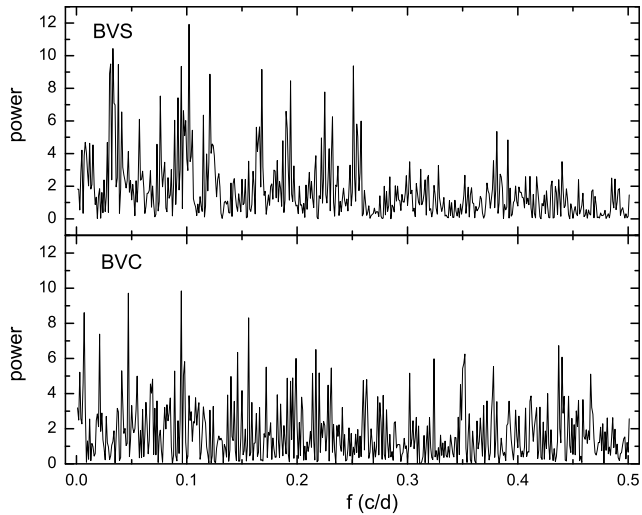}
\caption[]{The Lomb-Scargle periodogram of BVS (top) and BVC (bottom) of Ni I 6643.7}
\end{figure}

\subsection{Ca II H line Variation}

\begin{figure}
\label{CaII}
\centering
\includegraphics[width=9cm]{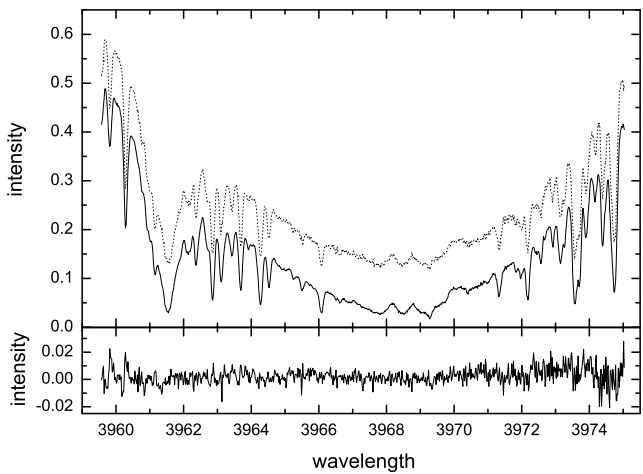}
\caption[]{Top: Plot of Ca II H line profile at negative RV phase (dotted line) and positive phase (solid line).
For the purpose of display, the dotted line is shifted upward by 0.1.
Bottom: The difference between the two line profiles.}
\end{figure}

Ca II H and K line profile variation, or S-Index was used to investigate the stellar activity and the nature of RV variation by several authors (Saar \& Donahue 1997, Queloz et al. 2001, Hatzes et al. 2003).
Unfortunately the S/N ratio of the spectra of BOES at the blue region is too low to compute the S-Index of each spectrum.
Instead of computing the S-Index of each spectrum,
we computed the Ca II H mean line profiles at the positive and negative RV phases to see any systematic difference between the two RV phases.
Figure 9 shows the computed line profiles. As we see in the plot, there is no systematic difference between the line profiles at the two RV phases.

\section{Discussion}
There is no doubt that a 429-d periodic variation is present in the RV measurements of {\gam1leo}.
This variation has persisted for almost 5 cycles with no change in phase or amplitude.
To clarify the cause of the periodic RV variation, we investigated the possibility of
any correlation between the RV variation and other stellar variations such as
Hipparcos photometry, EW of Ca II $\lambda$ 8662 line, line bisectors, and Ca II H line profile.
Within the accuracy of our measurements, we found no convincing evidence of such correlation.
Though we leave the real cause of the detected 429-d periodic RV variation open,
we conclude that the observed periodic RV variation is best explained by an unseen companion
with an estimated mass of m $\sin i = 8.78 \pm 0.2 M_{Jupiter}$.
The non-negligible eccentricity of 0.14 found from the orbital solution also lends some support
to a Keplerian orbital motion as the cause of the periodic RV variations.
But we would like to emphasize that continued RV observations of \gam1leo
are essential to clarify the true nature of the RV variation.
Perhaps the best way to confirm the orbital hypothesis is to detect the astrometric perturbation.
For \gam1leo the expected astrometric perturbation is several tenths of milli arcseconds,
which is about the limit of current ground-based astrometric techniques.

The nature of the 1340-d period found in the residual RVs remains unclear.
We do not see any evidence for this period in either Hipparcos photometry, Ca II EW, or line bisectors.
At face value this would argue for an additional companion, but this is not certain.
Continued RV measurements are needed to confirm a possible second companion.
We also found a significant power at P = 8.5 days in the residual RVs.
The 8.5-d period is close to the estimated fundamental radial pulsation period.
To confirm the reality of the 8.5-d period, more observations with better sampling are needed.

It is known that planet harboring giants does not show metal-rich tendency found from dwarf stars.
The low metalliicity of \gam1leo strengthens this trend further.
\gam1leo adds one more exoplanet discovered around a binary star system.
Since there are not still many exoplanet found around binary system,
we consider our discovery as a valuable addition to the field.

\acknowledgements
This work was supported by the Korea Foundation for International Cooperation of Science and Technology (KICOS)
through grant No. 07-179. DEM acknowledges his work as a part of research activity of the
Astrophysical Research Center for the Structure and Evolution of the Cosmos
(ARCSEC) which is supported by the Korean Science and Engineering Foundation.
We thank an anonymous referee for the great improvement of our paper.

{}

\Online
\begin{longtable}{ccc}
\caption{\label{rv} Relative RV measurements for \gam1leo by BOES}\\
\hline\hline
JD     & $\Delta$RV&  $\pm \sigma$    \\
day	   &   km\,s$^{-1}$ & km\,s$^{-1}$\\
\hline
\endfirsthead
\caption{continued.}\\
\hline\hline
JD    & $\Delta$RV   & $\pm \sigma$\\
day   & km\,s$^{-1}$ & km\,s$^{-1}$\\
\hline
\endhead
\hline
\endfoot
2452776.107091   &   0.3266 &    0.0208  \\
2452776.123872   &   0.3178 &    0.0080  \\
2452777.006125   &   0.2778 &    0.0091  \\
2452777.008682   &   0.2956 &    0.0081  \\
2452777.011761   &   0.2817 &    0.0070  \\
2452777.014538   &   0.2897 &    0.0075  \\
2452779.074177   &   0.2901 &    0.0077  \\
2452779.078169   &   0.2946 &    0.0075  \\
2452779.081873   &   0.2926 &    0.0083  \\
2452779.998369   &   0.2898 &    0.0082  \\
2452780.005556   &   0.2877 &    0.0079  \\
2452781.054575   &   0.2616 &    0.0080  \\
2452781.056889   &   0.2812 &    0.0081  \\
2452976.383870   &   0.0745 &    0.0067  \\
2452976.388523   &   0.0707 &    0.0064  \\
2452976.392621   &   0.0639 &    0.0072  \\
2452976.405643   &   0.0767 &    0.0076  \\
2452976.409497   &   0.0746 &    0.0063  \\
2452977.373912   &   0.0082 &    0.0076  \\
2452977.376826   &   0.0205 &    0.0077  \\
2452977.380842   &   0.0097 &    0.0055  \\
2452977.385148   &   0.0102 &    0.0064  \\
2452977.388725   &   0.0179 &    0.0061  \\
2452977.392278   &   0.0041 &    0.0065  \\
2452977.395890   &   0.0082 &    0.0064  \\
2453044.163185   &   0.2061 &    0.0080  \\
2453044.166761   &   0.2040 &    0.0071  \\
2453044.170731   &   0.2049 &    0.0073  \\
2453045.151698   &   0.2555 &    0.0060  \\
2453045.155518   &   0.2514 &    0.0054  \\
2453045.157763   &   0.2487 &    0.0053  \\
2453046.249436   &   0.2311 &    0.0066  \\
2453046.251925   &   0.2219 &    0.0063  \\
2453047.086196   &   0.2710 &    0.0072  \\
2453047.107528   &   0.2794 &    0.0083  \\
2453048.191778   &  0.2088  &   0.0067  \\
2453048.194174   &  0.2206  &   0.0065  \\
2453096.091559   &  0.3178  &   0.0059  \\
2453127.002214   &  0.2837  &   0.0061  \\
2453131.016505   &  0.3008  &   0.0064  \\
2453131.021805   &  0.2950  &   0.0059  \\
2453131.973915   &  0.3272  &   0.0063  \\
2453131.979319   &  0.3240  &   0.0054  \\
2453132.981655   &  0.3177  &   0.0064  \\
2453132.985162   &  0.3178  &   0.0057  \\
2453354.323803   &  0.0774  &   0.0061  \\
2453430.116894   &  0.2661  &   0.0069  \\
2453459.072528   &  0.2747  &   0.0057  \\
2453459.159942   &  0.2675  &   0.0058  \\
2453460.098807   &  0.2919  &   0.0063  \\
2453460.154093   &  0.2953  &   0.0058  \\
2453460.195850   &  0.3012  &   0.0057  \\
2453510.037239   &  0.3110  &   0.0060  \\
2453511.032443   &  0.3961  &   0.0051  \\
2453511.036193   &  0.4019  &   0.0051  \\
2453778.295467   &  0.0157  &   0.0059  \\
2453778.301347   &  0.0173  &   0.0058  \\
2453779.201277   &  0.0194  &   0.0064  \\
2453817.101751   & -0.0045  &   0.0051  \\
2453819.133068   &  0.0368  &   0.0046  \\
2453820.047891   & -0.0106  &   0.0055  \\
2453869.044301   &  0.2395  &   0.0057  \\
2453888.036739   &  0.2663  &   0.0058  \\
2453896.006225   &  0.2613  &   0.0053  \\
2454038.363098   &  0.3835  &   0.0048  \\
2454081.293115   &  0.2434  &   0.0056  \\
2454123.162010   &  0.1156  &   0.0040  \\
2454123.165055   &  0.1182  &   0.0047  \\
2454125.186280   &  0.1612  &   0.0048  \\
2454125.189092   &  0.1663  &   0.0046  \\
2454147.244498   &  0.0871  &   0.0052  \\
2454147.246338   &  0.0835  &   0.0052  \\
2454151.307380   &  0.0405  &   0.0052  \\
2454151.311002   &  0.0372  &   0.0056  \\
2454205.151851   & -0.0403  &   0.0101  \\
2454207.097911   & -0.0063  &   0.0067  \\
2454209.085063   & -0.0704  &   0.0067  \\
2454209.088501   & -0.0672  &   0.0064  \\
2454210.138147   & -0.0527  &   0.0064  \\
2454214.070984   & -0.0761  &   0.0101  \\
2454223.050863   & -0.0905  &   0.0063  \\
2454223.057657   & -0.0883  &   0.0063  \\
2454223.060886   & -0.0876  &   0.0057  \\
2454223.064265   & -0.0845  &   0.0057  \\
2454224.020820   & -0.0226  &   0.0080  \\
2454224.026977   & -0.0231  &   0.0061  \\
2454224.034580   & -0.0210  &   0.0058  \\
2454224.042022   & -0.0199  &   0.0059  \\
2454225.049231   & -0.0266  &   0.0054  \\
2454225.052067   & -0.0292  &   0.0054  \\
2454225.054694   & -0.0262  &   0.0053  \\
2454225.057332   & -0.0276  &   0.0054  \\
2454227.987093   & -0.0323  &   0.0050  \\
2454227.990125   & -0.0319  &   0.0052  \\
2454227.992289   & -0.0358  &   0.0053  \\
2454227.994059   & -0.0301  &   0.0055  \\
2454228.961109   & -0.0871  &   0.0060  \\
2454228.963111   & -0.0839  &   0.0057  \\
2454228.965507   & -0.0854  &   0.0052  \\
2454228.967243   & -0.0875  &   0.0052  \\
2454230.968847   & -0.0619  &   0.0058  \\
2454230.970386   & -0.0725  &   0.0055  \\
2454230.972156   & -0.0655  &   0.0054  \\
2454230.973904   & -0.0677  &   0.0055  \\
2454235.999486   & -0.0195  &   0.0065  \\
2454236.002368   &  -0.0218 &    0.0071  \\
2454236.005226   &  -0.0261 &    0.0070  \\
2454241.994669   &  -0.0246 &    0.0050  \\
2454241.997296   &  -0.0228 &    0.0052  \\
2454241.999148   &  -0.0269 &    0.0050  \\
2454242.001023   &  -0.0229 &    0.0058  \\
2454242.002944   &  -0.0217 &    0.0049  \\
2454242.963823   &  -0.0113 &    0.0055  \\
2454242.965848   &  -0.0071 &    0.0054  \\
2454254.986465   &  -0.0289 &    0.0063  \\
2454254.989184   &  -0.0256 &    0.0062  \\
2454263.003267   &  -0.0104 &    0.0090  \\
2454263.007352   &  -0.0257 &    0.0095  \\
2454263.013960   &  -0.0163 &    0.0077  \\
2454264.031125   &   0.0325 &    0.0075  \\
2454264.033671   &   0.0313 &    0.0066  \\
2454264.036889   &   0.0324 &    0.0068  \\
2454269.016633   &   0.0463 &    0.0089  \\
2454269.018889   &   0.0527 &    0.0074  \\
2454269.021436   &   0.0479 &    0.0080  \\
2454273.989254   &   0.0845 &    0.0062  \\
2454273.992078   &   0.0908 &    0.0063  \\
2454273.995110   &   0.0781 &    0.0060  \\
2454388.351691   &   0.3225 &    0.0068  \\
2454388.355730   &   0.3202 &    0.0074  \\
2454389.372500   &   0.3566 &    0.0063  \\
2454389.376540   &   0.3543 &    0.0057  \\
2454389.380649   &   0.3559 &    0.0064  \\
2454390.370681   &   0.2859 &    0.0057  \\
2454390.374732   &   0.2907 &    0.0059  \\
2454396.375832   &   0.3342 &    0.0073  \\
2454396.379687   &   0.3378 &    0.0064  \\
2454426.380085   &   0.3521 &    0.0063  \\
2454426.381393   &   0.3492 &    0.0060  \\
2454426.382551   &   0.3547 &    0.0065  \\
2454452.404499   &   0.3826 &    0.0057  \\
2454452.406304   &   0.3888 &    0.0067  \\
2454458.354959   &   0.2805 &    0.0068  \\
2454458.357147   &   0.2849 &    0.0068  \\
2454470.216053   &   0.3360 &    0.0059  \\
2454470.219259   &   0.3257 &    0.0062  \\
2454470.222465   &   0.3287 &    0.0062  \\
2454470.225671   &   0.3390 &    0.0064  \\
2454472.224505   &   0.3220 &    0.0058  \\
2454472.227190   &   0.3172 &    0.0093  \\
2454472.229875   &   0.3226 &    0.0057  \\
2454472.233429   &   0.3080 &    0.0090  \\
2454483.168020   &   0.2081 &    0.0054  \\
2454483.170358   &   0.2109 &    0.0055  \\
2454483.172175   &   0.2149 &    0.0054  \\
2454483.174340   &   0.2167 &    0.0055  \\
2454505.177577   &   0.2174 &    0.0082  \\
2454506.201658   &   0.2736 &    0.0060  \\
2454506.203452   &   0.2765 &    0.0060  \\
2454506.205327   &   0.2779 &    0.0056  \\
2454506.207653   &   0.2703 &    0.0063  \\
2454516.250170   &   0.2313 &    0.0058  \\
2454516.253353   &   0.2294 &    0.0062  \\
2454536.133769   &   0.1511 &    0.0057  \\
2454536.135667   &   0.1507 &    0.0058  \\
2454536.137437   &   0.1569 &    0.0060  \\
2454536.139000   &   0.1524 &    0.0057  \\
2454537.181626   &   0.2148 &    0.0068  \\
2454537.184138   &   0.2027 &    0.0074  \\
2454537.187795   &   0.2130 &    0.0059  \\
2454537.191799   &   0.2086 &    0.0064  \\
2454549.967591   &   0.1207 &    0.0063  \\
2454549.970403   &   0.1211 &    0.0058  \\
2454550.959573   &   0.1428 &    0.0053  \\
2454550.962675   &   0.1390 &    0.0059  \\
2454550.966297   &   0.1427 &    0.0054  \\
2454551.954656   &   0.0641 &    0.0082  \\
2454551.957538   &   0.0720 &    0.0069  \\
2454551.960744   &   0.0711 &    0.0072  \\
2454551.963672   &   0.0746 &    0.0067  \\
2454551.967016   &   0.0659 &    0.0066  \\
2454551.969944   &   0.0639 &    0.0068  \\
2454575.020263   &   0.0246 &    0.0061  \\
2454575.022323   &   0.0189 &    0.0054  \\
2454575.024556   &   0.0273 &    0.0060  \\
2454575.026882   &   0.0177 &    0.0058  \\
2454618.043283   &  -0.0143 &    0.0055  \\
2454618.045169   &  -0.0119 &    0.0062  \\
2454618.047056   &  -0.0083 &    0.0059  \\
2454618.048942   &  -0.0107 &    0.0061  \\
2454618.997301   &  -0.0566 &    0.0071  \\
2454619.002659   &  -0.0604 &    0.0067  \\
2454834.279749   &   0.4557 &    0.0050  \\
2454834.281566   &   0.4511 &    0.0046  \\
2454847.269458   &   0.3762 &    0.0054  \\
2454847.271253   &   0.3742 &    0.0052  \\
2454847.273047   &   0.3765 &    0.0051  \\
2454847.274841   &   0.3724 &    0.0055  \\
2454879.093335   &   0.4144 &    0.0055  \\
2454879.095950   &   0.4090 &    0.0057  \\
2454929.088055   &   0.3674 &    0.0122  \\
2454929.093321   &   0.3762 &    0.0099  \\
2454930.056224   &   0.3517 &    0.0067  \\
2454930.058573   &   0.3466 &    0.0060  \\
2454971.053095   &   0.2712 &    0.0065  \\
\end{longtable}
\end{document}